\documentstyle[twoside,psfig]{article}

\catcode`\@=11
\long\def\@makefntext#1{
\protect\noindent \hbox to 3.2pt {\hskip-.9pt  
$^{{\eightrm\@thefnmark}}$\hfil}#1\hfill}		

\def\@makefnmark{\hbox to 0pt{$^{\@thefnmark}$\hss}}	
	
\def\ps@myheadings{\let\@mkboth\@gobbletwo
\def\@oddhead{\hbox{}
\rightmark\hfil\eightrm\thepage}   
\def\@oddfoot{}\def\@evenhead{\eightrm\thepage\hfil
\leftmark\hbox{}}\def\@evenfoot{}
\def\sectionmark##1{}\def\subsectionmark##1{}}

\oddsidemargin=\evensidemargin
\newcounter{sectionc}\newcounter{subsectionc}\newcounter{subsubsectionc}
\renewcommand{\section}[1] {\vspace{12pt}\addtocounter{sectionc}{1} 
\setcounter{subsectionc}{0}\setcounter{subsubsectionc}{0}\noindent 
	{\tenbf\thesectionc. #1}\par\vspace{5pt}}
\renewcommand{\subsection}[1] {\vspace{12pt}\addtocounter{subsectionc}{1} 
	\setcounter{subsubsectionc}{0}\noindent 
	{\bf\thesectionc.\thesubsectionc. {\kern1pt \bfit #1}}\par\vspace{5pt}}
\renewcommand{\subsubsection}[1] {\vspace{12pt}\addtocounter{subsubsectionc}{1}
	\noindent{\tenrm\thesectionc.\thesubsectionc.\thesubsubsectionc.
	{\kern1pt \tenit #1}}\par\vspace{5pt}}
\newcommand{\nonumsection}[1] {\vspace{12pt}\noindent{\tenbf #1}
	\par\vspace{5pt}}

\newcounter{appendixc}
\newcounter{subappendixc}[appendixc]
\newcounter{subsubappendixc}[subappendixc]
\renewcommand{\thesubappendixc}{\Alph{appendixc}.\arabic{subappendixc}}
\renewcommand{\thesubsubappendixc}
	{\Alph{appendixc}.\arabic{subappendixc}.\arabic{subsubappendixc}}

\renewcommand{\appendix}[1] {\vspace{12pt}
        \refstepcounter{appendixc}
        \setcounter{figure}{0}
        \setcounter{table}{0}
        \setcounter{lemma}{0}
        \setcounter{theorem}{0}
        \setcounter{corollary}{0}
        \setcounter{definition}{0}
        \setcounter{equation}{0}
        \renewcommand{\thefigure}{\Alph{appendixc}.\arabic{figure}}
        \renewcommand{\thetable}{\Alph{appendixc}.\arabic{table}}
        \renewcommand{\theappendixc}{\Alph{appendixc}}
        \renewcommand{\thelemma}{\Alph{appendixc}.\arabic{lemma}}
        \renewcommand{\thetheorem}{\Alph{appendixc}.\arabic{theorem}}
        \renewcommand{\thedefinition}{\Alph{appendixc}.\arabic{definition}}
        \renewcommand{\thecorollary}{\Alph{appendixc}.\arabic{corollary}}
        \renewcommand{\theequation}{\Alph{appendixc}.\arabic{equation}}
        \noindent{\tenbf Appendix \theappendixc #1}\par\vspace{5pt}}
\newcommand{\subappendix}[1] {\vspace{12pt}
        \refstepcounter{subappendixc}
        \noindent{\bf Appendix \thesubappendixc. {\kern1pt \bfit #1}}
	\par\vspace{5pt}}
\newcommand{\subsubappendix}[1] {\vspace{12pt}
        \refstepcounter{subsubappendixc}
        \noindent{\rm Appendix \thesubsubappendixc. {\kern1pt \tenit #1}}
	\par\vspace{5pt}}

\newcommand{\textlineskip}{\baselineskip=13pt}
\newcommand{\smalllineskip}{\baselineskip=10pt}

\def\eightcirc{
\begin{picture}(0,0)
\put(4.4,1.8){\circle{6.5}}
\end{picture}}
\def\eightcopyright{\eightcirc\kern2.7pt\hbox{\eightrm c}} 

\newcommand{\copyrightheading}[1]
	{\vspace*{-2.5cm}\smalllineskip{\flushleft
	{\footnotesize International Journal of Modern Physics A, #1}\\
	{\footnotesize $\eightcopyright$\, World Scientific Publishing
	 Company}\\
	 }}

\def\abstracts#1#2#3{{
	\centering{\begin{minipage}{4.5in}\baselineskip=10pt\footnotesize
	\parindent=0pt #1\par 
	\parindent=15pt #2\par
	\parindent=15pt #3
	\end{minipage}}\par}} 

\renewenvironment{thebibliography}[1]
	{\frenchspacing
	 \ninerm\baselineskip=11pt
	 \begin{list}{\arabic{enumi}.}
	{\usecounter{enumi}\setlength{\parsep}{0pt}
	 \setlength{\leftmargin 12.7pt}{\rightmargin 0pt} 
	 \setlength{\itemsep}{0pt} \settowidth
	{\labelwidth}{#1.}\sloppy}}{\end{list}}

\newcounter{itemlistc}
\newcounter{romanlistc}
\newcounter{alphlistc}
\newcounter{arabiclistc}

\newcommand{\fcaption}[1]{
        \refstepcounter{figure}
        \setbox\@tempboxa = \hbox{\footnotesize Fig.~\thefigure. #1}
        \ifdim \wd\@tempboxa > 5in
           {\begin{center}
        \parbox{5in}{\footnotesize\smalllineskip Fig.~\thefigure. #1}
            \end{center}}
        \else
             {\begin{center}
             {\footnotesize Fig.~\thefigure. #1}
              \end{center}}
        \fi}

\newcommand{\tcaption}[1]{
        \refstepcounter{table}
        \setbox\@tempboxa = \hbox{\footnotesize Table~\thetable. #1}
        \ifdim \wd\@tempboxa > 5in
           {\begin{center}
        \parbox{5in}{\footnotesize\smalllineskip Table~\thetable. #1}
            \end{center}}
        \else
             {\begin{center}
             {\footnotesize Table~\thetable. #1}
              \end{center}}
        \fi}

\def\@citex[#1]#2{\if@filesw\immediate\write\@auxout
	{\string\citation{#2}}\fi
\def\@citea{}\@cite{\@for\@citeb:=#2\do
	{\@citea\def\@citea{,}\@ifundefined
	{b@\@citeb}{{\bf ?}\@warning
	{Citation `\@citeb' on page \thepage \space undefined}}
	{\csname b@\@citeb\endcsname}}}{#1}}

\newif\if@cghi
\def\cite{\@cghitrue\@ifnextchar [{\@tempswatrue
	\@citex}{\@tempswafalse\@citex[]}}
\def\citelow{\@cghifalse\@ifnextchar [{\@tempswatrue
	\@citex}{\@tempswafalse\@citex[]}}
\def\@cite#1#2{{$\null^{#1}$\if@tempswa\typeout
	{IJCGA warning: optional citation argument 
	ignored: `#2'} \fi}}

\def\pmb#1{\setbox0=\hbox{#1}
	\kern-.025em\copy0\kern-\wd0
	\kern.05em\copy0\kern-\wd0
	\kern-.025em\raise.0433em\box0}

\def\fnt#1#2{\footnotetext{\kern-.3em
	{$^{\mbox{\scriptsize #1}}$}{#2}}}

\def\fpage#1{\begingroup
\voffset=.3in
\thispagestyle{empty}\begin{table}[b]\centerline{\footnotesize #1}
	\end{table}\endgroup}

\def\runninghead#1#2{\pagestyle{myheadings}
\markboth{{\protect\footnotesize\it{\quad #1}}\hfill}
{\hfill{\protect\footnotesize\it{#2\quad}}}}
\font\tenrm=cmr10
\font\tenit=cmti10 
\font\tenbf=cmbx10
\font\bfit=cmbxti10 at 10pt
\font\ninerm=cmr9

\font\eightrm=cmr8

\def\qed{\hbox{${\vcenter{\vbox{			
   \hrule height 0.4pt\hbox{\vrule width 0.4pt height 6pt
   \kern5pt\vrule width 0.4pt}\hrule height 0.4pt}}}$}}

\def \missing {$E_{T} \mbox{\hspace{-0.42cm}}/ \mbox{\hspace{0.31cm}}$}
\def \missed  {$E_{T} \mbox{\hspace{-0.43cm}}/ \mbox{\hspace{0.4cm}}$}
\def \misspar {$E_{T} \mbox{\hspace{-0.42cm}}/ \mbox{\hspace{0.22cm}}$}

\def \missab  {$E_{T} \mbox{\hspace{-0.38cm}}/ \mbox{\hspace{0.31cm}}$}
\def \missabp {$E_{T} \mbox{\hspace{-0.38cm}}/ \mbox{\hspace{0.22cm}}$}

\def \metmin  {$E_{T}^{min} \mbox{\hspace{-0.76cm}}/ \mbox{\hspace{0.63cm}}$}

\begin{document}

\runninghead{Search for SUSY with \missab and Jets at CDF} 
            {Richard Haas, University of Florida}

\normalsize\textlineskip
\thispagestyle{empty}
\setcounter{page}{1}

\copyrightheading{}			


\vspace*{0.88truein}

\fpage{1}
\centerline{\bf SEARCH FOR SUSY WITH \missed AND JETS AT CDF}
\vspace*{0.37truein}
\centerline{\footnotesize RICHARD HAAS\footnote{rhaas@cdfsga.fnal.gov}}
\vspace*{0.015truein}
\centerline{\footnotesize (Representing the CDF Collaboration)}
\vspace*{0.015truein}
\centerline{\footnotesize\it Department of Physics, University of Florida}
\baselineskip=10pt
\centerline{\footnotesize\it Gainesville, FL 32611, USA}

\vspace*{0.21truein}
\abstracts{
   Events with signatures involving large missing transverse energy 
   (\missabp) are among the 
   quintessential search modes for R-parity conserving supersymmetry.
   CDF has conducted two recent analyses for supersymmetry which
   use \missab and jets.
   The \missab and monojet signature is employed to determine process
   independent limits for the production of new physics beyond the
   Standard Model and then applied to models of spontaneous breaking
   of supersymmetry to determine limits on the
   supersymmetry breaking parameter and the gravitino mass. 
   Direct searches for scalar top and scalar bottom quarks within the
   framework of supersymmetric models are performed using a signature
   of \missab and two heavy flavor jets.
   Since the data is found to be consistent with Standard Model
   expectations, limits are determined in the mass planes 
   $m (\tilde{\chi}_{1}^{0}) - m (\tilde{t}_{1})$ and
   $m (\tilde{\chi}_{1}^{0}) - m (\tilde{b}_{1})$.
}{}{}

\vspace*{1pt}\textlineskip
\section{New Physics and Gravitinos}
\vspace*{-0.5pt}
\noindent


Several theories beyond the Standard Model\cite{mettheory}
contain light, neutral particles which have extremely small
interaction cross sections.
These particles typically pass through detectors unobserved 
producing a signature with an imbalance in transverse energy 
($E_{T}$) quantified as missing transverse energy (\misspar).

\begin{figure}
   \vspace*{13pt}
   \centerline{\psfig{file=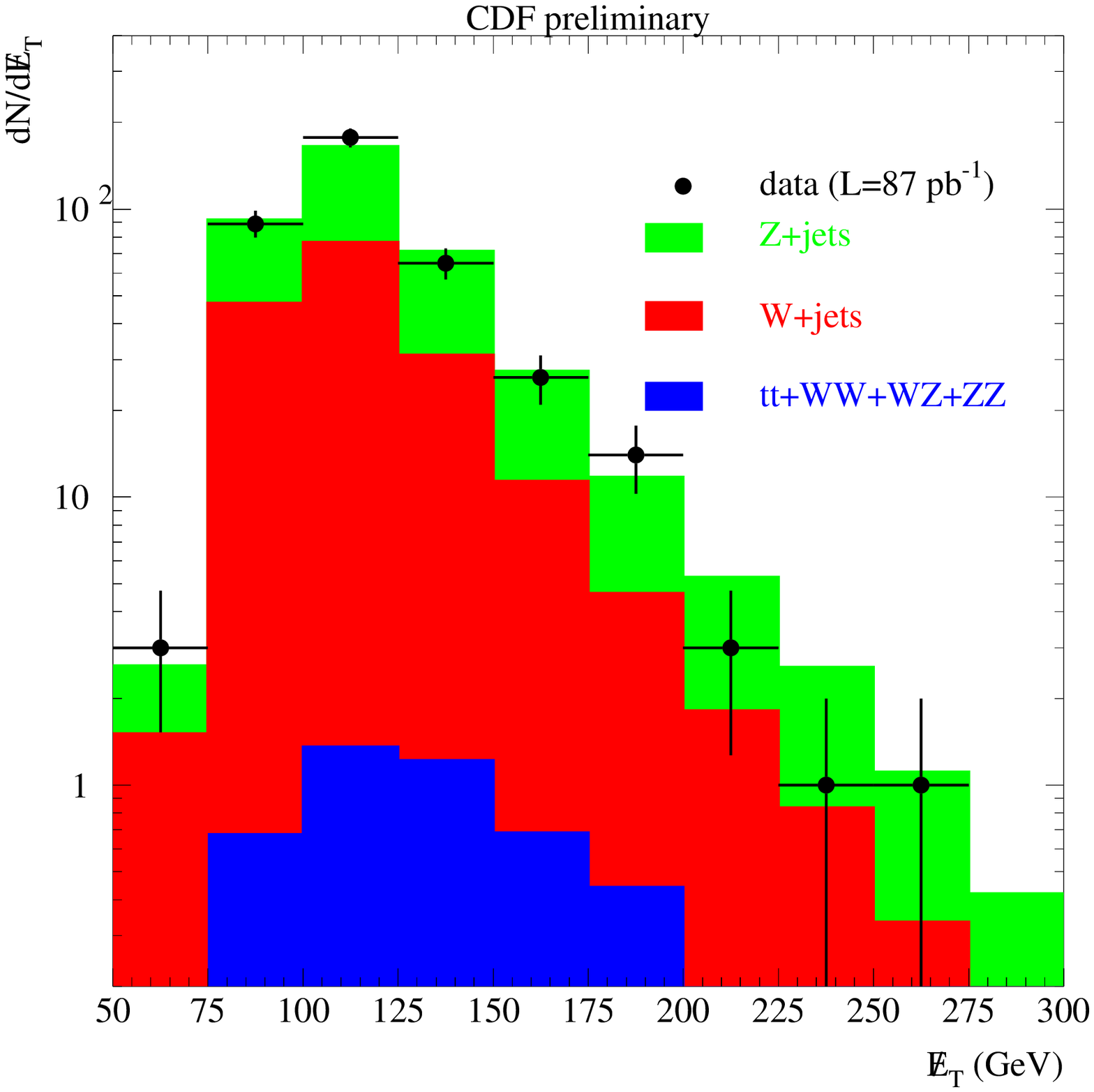,width=2.4in}
               \hspace{0.2in}
               \psfig{file=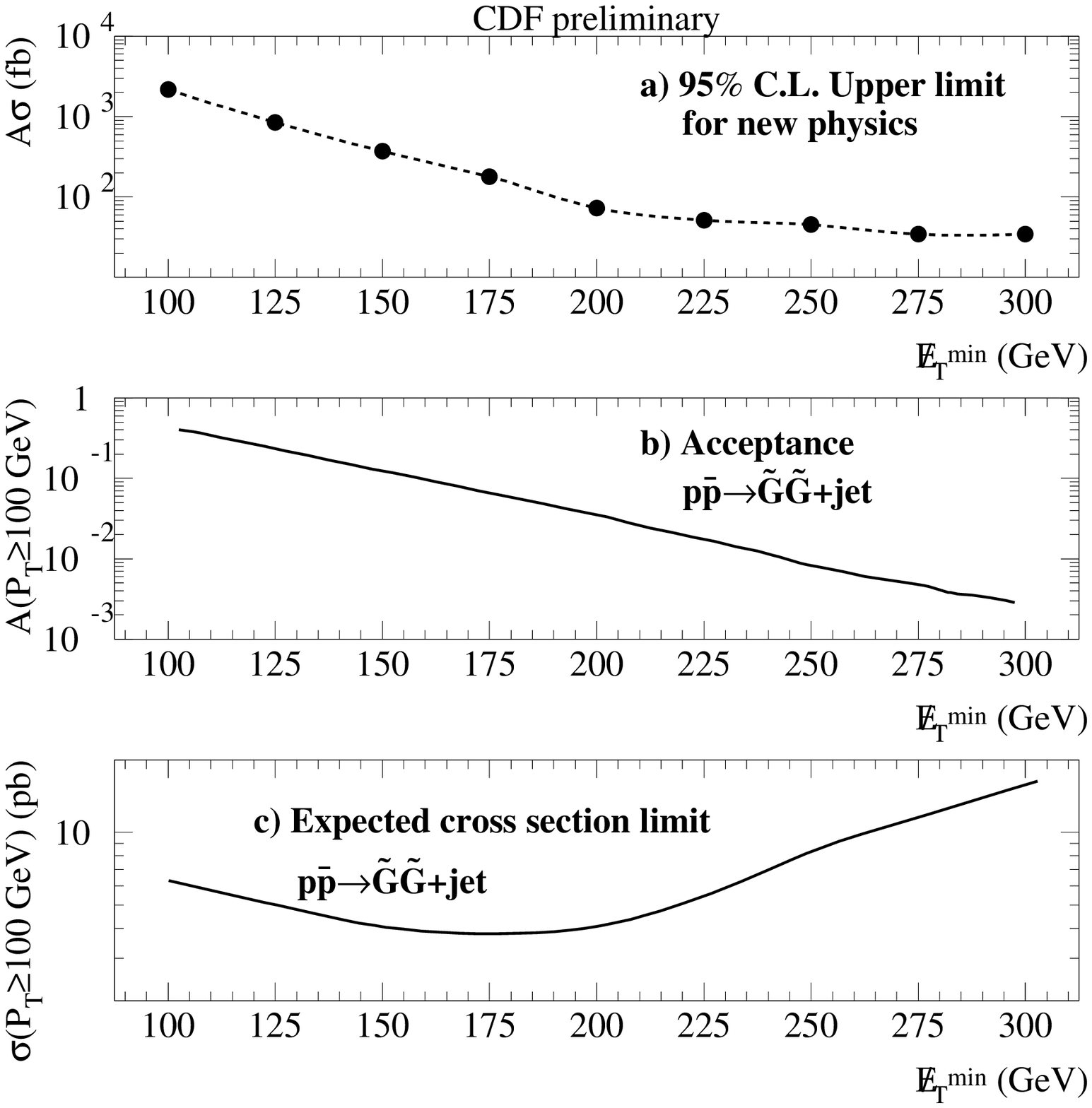,width=2.4in}}
   \vspace*{13pt}
   \fcaption{Left plot: Distribution of \missing for data compared to
             background expectations. 
             Right plot: 95\% CL upper limit for acceptance times
             production cross section of new physics (top), 
             the signal acceptance for $\tilde{G} \tilde{G}$ jet
             processes with 
             $p_{T}^{\tilde{G} \tilde{G}} \geq 100$ GeV/c (middle), 
             and the expected 95\% CL upper limit for the cross section 
             of $\tilde{G} \tilde{G}$ jet production with 
             $p_{T}^{\tilde{G} \tilde{G}} \geq 100$ GeV/c (bottom).
            }
   \label{fig:gravitino}
\end{figure}

CDF has performed a search for physics beyond the Standard Model 
by examining events with 
large \missing and a high $E_{T}$ jet.\cite{gravitino}
Trigger induced systematic effects are reduced by increasing the
$35$ GeV trigger threshold to \missing $\geq 50$ GeV. 
A high $E_{T}$ jet is obtained by selecting events which contain 
a leading jet with $E_{T} \geq 80$ GeV. 
The effects of jet energy mismeasurement and additional
instrumental backgrounds are reduced by requiring that the angle
between the \missing direction and the nearest jet 
satisfy $\Delta \phi ($\missing , jet$) \geq 90^{\circ}$. 
Events with identified electrons or muons are removed.
The \missing spectrum and the estimated backgrounds after 
all selection criteria have been applied are shown in the left plot of
Figure \ref{fig:gravitino}. The number of observed events is 379 while
$380 \pm 129$ background events are determined.
The 95\% confidence level (CL) upper limit for the product of 
acceptance and cross section for new physics as a function of \metmin ,
the \missing threshold, are shown in the top right
plot of Figure \ref{fig:gravitino}. 
To determine a process specific cross section limit, 
the pertinent acceptance is calculated.


Theories in which supersymmetry is broken via standard model gauge
interactions can produce signatures with large \missing and high 
$E_{T}$ jets. 
In models with gravity where supersymmetry is realized locally,  
the gravitino ($\tilde{G}$) acquires a mass
$m_{\tilde{G}}$.\cite{rattazzi} 
A gravitino with mass much less than the electroweak scale may be 
created when the scale at which the breaking of supersymmetry is 
transmitted is much less than the Planck scale. If R-parity is
conserved, this naturally provides for a gravitino which can be
the LSP. 
If sufficiently light ($m_{\tilde{G}} \ll 10^{-4}$ eV/c$^{2}$), the
gravitino may appear at the Tevatron in signatures characterized by 
large \missing .


Pair production of gravitinos in conjunction with jets may occur
through the processes 
$q \bar{q} \rightarrow \tilde{G} \tilde{G} g$, 
$qg \rightarrow \tilde{G} \tilde{G} q$, 
$\bar{q} g \rightarrow \tilde{G} \tilde{G} \bar{q}$, and
$gg \rightarrow \tilde{G} \tilde{G} g$. The resulting signature for
such events would be large \missing and a jet.\cite{gravitino}
In order to determine
acceptances for $\tilde{G} \tilde{G}$ jet production, the theoretical
predications\cite{grav_theory} calculated at the supersymmetry
breaking scale $\sqrt{F} = 200$ GeV 
are employed together with a CDF detector simulation.
The dependence on \metmin in the calculated acceptances and cross
sections is minimized by specifying 
the value of the transverse magnitude of the vector sum of
the two gravitino momenta before further radiation,
$p_{T}^{\tilde{G} \tilde{G}}$. 
For the acceptances and cross sections, 
$p_{T}^{\tilde{G} \tilde{G}} \geq 100$ GeV/c. The signal acceptance
resulting from the simulations is shown in the middle right plot of
Figure \ref{fig:gravitino}.


The expected 95\% CL cross section upper limit for 
$\tilde{G} \tilde{G}$ jet production as a function of \metmin is 
shown in the bottom right plot of Figure \ref{fig:gravitino}. 
The value of \metmin is selected which results in the optimal reach in
cross section. For \metmin $= 175$ GeV, 19 events are observed with an
estimated background of $21.6 \pm 7.0$. Incorporating a 20\%
uncertainty from the acceptance and a 4\% uncertainty from the
luminosity, the 95\% CL upper limit for the number of signal events is
$16.9$. With an acceptance of $6.2$\%, the 95\% CL upper limit for the
$\tilde{G} \tilde{G}$ jet production cross section with
$p_{T}^{\tilde{G} \tilde{G}} \geq 100$ GeV/c is $3.1$ pb.
Together with the cross section calculated from theory of 
$12.6 \pm 4.0$ pb for $\sqrt{F} = 200$ GeV with a 32\% systematic 
uncertainty, the 95\% CL limit on the 
supersymmetry breaking scale is determined to be  
$\sqrt{F} \geq 217$ GeV. This value for $\sqrt{F}$ corresponds to
gravitino masses residing above $1.1 \times 10^{-5}$ eV/c$^{2}$.

\vspace*{1pt}\textlineskip
\section{Supersymmetric Top and Bottom}
\vspace*{-0.5pt}
\noindent


Theoretical considerations suggest that the scalar top quark
($\tilde{t}$) may be the lightest scalar quark and potentially lighter
than the top quark. 
The dependence of the splitting in the $\tilde{t}$ mass eigenvalues
on the top quark mass due to the mixing of chiral scalar top states 
can cause widely disparate values allowing $\tilde{t}_{1}$ to be much 
lighter than $\tilde{t}_{2}$.
The values which anchor the mass spectrum are determined by scaling of
the renormalization group equations from high energies 
($\sim M_{\mbox{\scriptsize p}}$) down to the electroweak regime 
($\sim 1$ TeV). 
The effect on the renormalization group equations of the top Yukawa
coupling is to drive the masses of the scalar top quarks lower than the
first two families.

Although the splitting between $\tilde{b}$ mass eigenvalues due to the
bottom mass and the influence of the bottom Yukawa coupling are
weaker, values of supersymmetric parameters exist which cause large 
mixing among the scalar bottom quarks. 
Through a judicious choice of the ratio of the vacuum expectation
values of the Higgs fields, $\tan \beta$, large mixing between the
states can occur leading to considerable splitting among the
$\tilde{b}$ mass eigenvalues and a small mass for the lighter mass 
eigenstate. This condition can appear in the $\tilde{b}$ sector 
when $\tan \beta$ is large.


At the Tevatron, the principal mechanisms for pair production of third
generation scalar quarks are $q \bar{q}$ annihilation and gluon
fusion. Several potential decay modes exist for 
$\tilde{t}_{1}$.\cite{stopdecays} 
The process $\tilde{t}_{1} \rightarrow c \tilde{\chi}_{1}^{0}$ 
occurring via Feynman diagram loops and ordinarily suppressed dominates 
if 
$m_{\tilde{t}_{1}} < m_{b} + m_{\tilde{\chi}_{1}^{\pm}}$ and 
$m_{\tilde{t}_{1}} < m_{b} + m_{\ell} + m_{\tilde{\nu}}$.
For $\tilde{b}$, if $m_{t}$, 
$m_{\tilde{\chi}_{1}^{+}} > m_{\tilde{b}} > m_{b} + m_{\tilde{\chi}_{1}^{0}}$,
the decay mode $\tilde{b} \rightarrow b \tilde{\chi}^{0}_{1}$ 
is open and dominant.


A signature of 
two heavy flavor jets, missing transverse energy, and the
absence of leptons is utilized at CDF to search for pair produced 
scalar top and scalar bottom quarks decaying to 
$c \bar{c} \tilde{\chi}_{1}^{0} \bar{\tilde{\chi}}_{1}^{0}$ and 
$b \bar{b} \tilde{\chi}_{1}^{0} \bar{\tilde{\chi}}_{1}^{0}$,
respectively.\cite{stopsbottom} 
Events characterized by two or three hard jets with 
$E_{T} \geq 15$ GeV and $| \eta | \leq 2$ and no additional jets 
with $E_{T} > 7$ GeV and $| \eta | \leq 3.6$ are selected.
Systematic effects caused by the \missing trigger threshold 
are suppressed by increasing the \missing requirement from $35$ GeV 
to $40$ GeV. 
The effects of jet energy mismeasurement and QCD background are
reduced by ensuring that the jets are well separated from the 
direction of \missing as well as from each other. 
The angles are restricted between 
\missing and any jet ($\Delta \phi($\missing$, j) > 45^{\circ}$), 
\missing and the leading jet 
($\Delta \phi($\missing$,j_{1}) < 165^{\circ}$), and
the two highest $E_{T}$ jets 
($45^{\circ} < \Delta \phi(j_{1}, j_{2}) < 165^{\circ}$).
Events containing electrons and muons are rejected. 
Jets arising from $c$ and $b$ quarks are identified
using the jet probability algorithm\cite{jetprob} which gives the
probability, ${\cal P}_{\mbox{\scriptsize jet}}$, 
that the ensemble of tracks in a jet is consistent with originating 
from a primary vertex. 
For events corresponding to $\tilde{t}$ decays, 
${\cal P}_{\mbox{\scriptsize jet}} \leq 0.05$. For $\tilde{b}$ decays,
${\cal P}_{\mbox{\scriptsize jet}} \leq 0.01$.


\begin{figure}
   \vspace*{13pt}
   \centerline{\psfig{file=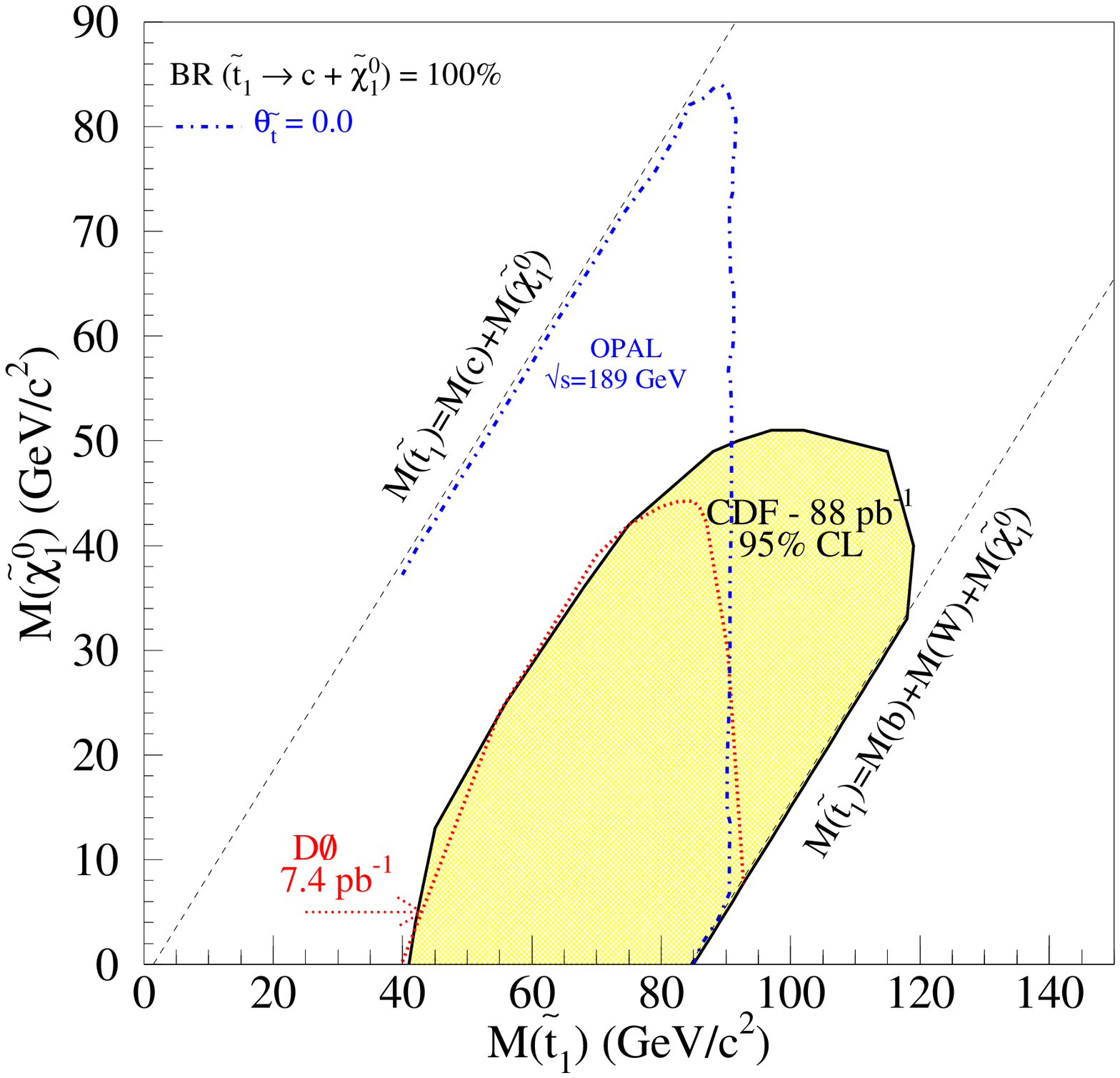,width=2.4in}
               \hspace{0.2in}
               \psfig{file=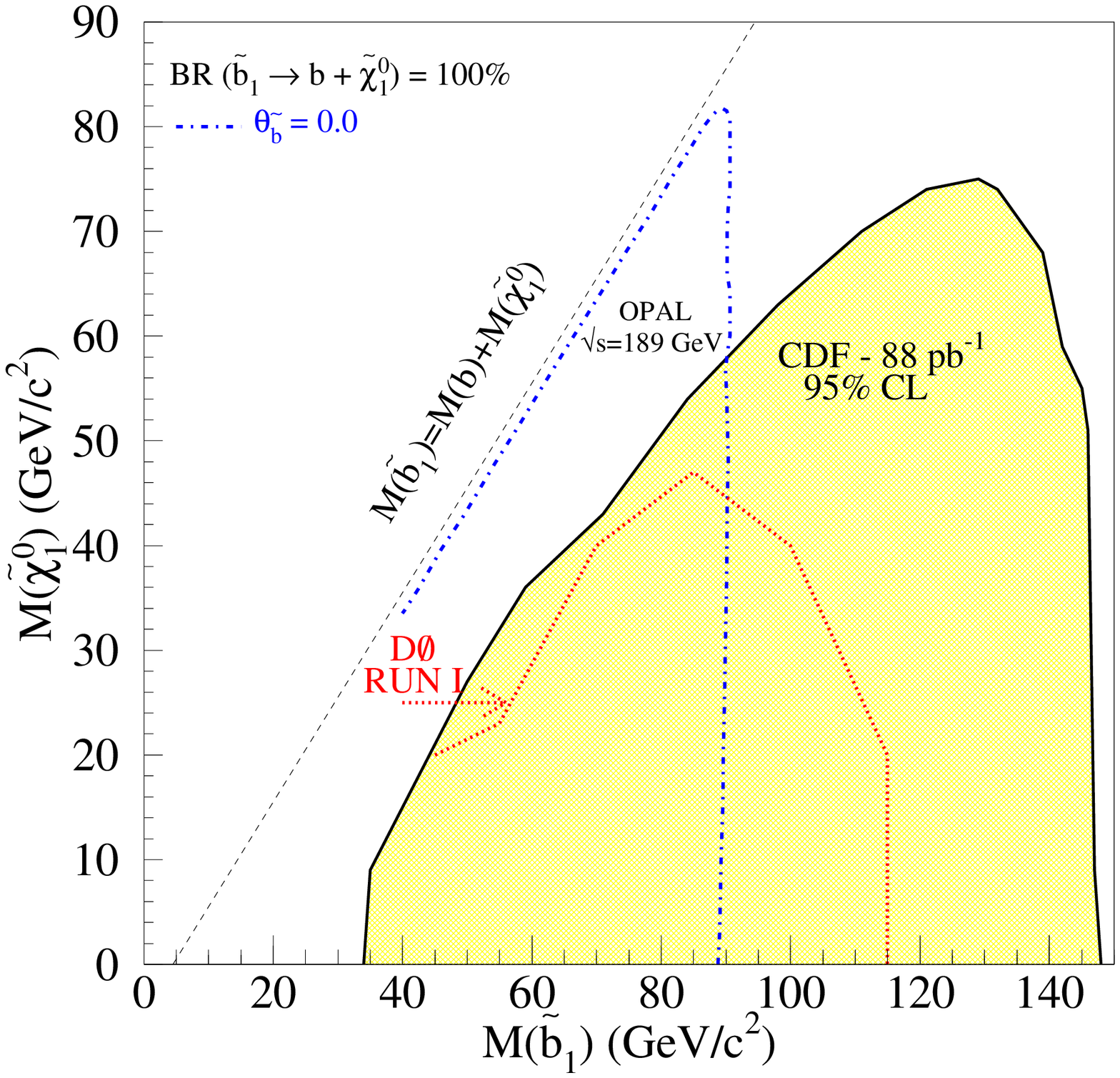,width=2.4in}}
   \vspace*{13pt}
   \fcaption{Left plot: CDF 95\% CL exclusion region in 
             $m (\tilde{\chi}_{1}^{0}) - m (\tilde{t}_{1})$ plane for
             $\tilde{t}_{1} \rightarrow c \tilde{\chi}_{1}^{0}$.
             Right plot: CDF 95\% CL exclusion region in 
             $m (\tilde{\chi}_{1}^{0}) - m (\tilde{b}_{1})$ plane for 
             $\tilde{b}_{1} \rightarrow b_{1} \tilde{\chi}_{1}^{0}$.
            }
   \label{fig:susy}
\end{figure}

For the scalar top quarks search, 11 observed events are found and a
background of $14.5 \pm 4.2$ is estimated. Since there is no excess 
of observed events over Standard Model backgrounds, 
a 95\% CL exclusion region in the 
$m (\tilde{\chi}_{1}^{0}) - m (\tilde{t}_{1})$ plane is
determined using a background subtraction method.\cite{back}
The results are shown in the left plot of Figure \ref{fig:susy}.
The break in sensitivity between the 95\% CL excluded region and the
kinematic limit is primarily due to requiring \missing $\geq 40$ GeV. 
However, the recent analysis performed by OPAL\cite{opal} without a
\missing restriction supplement the coverage obtained by CDF and are
shown in the figure.
For $m (\tilde{\chi}_{1}^{0}) = 40$ GeV/c$^{2}$, the maximum scalar
top quark mass excluded is 119 GeV/c$^{2}$ at the 95\% CL.

Once the selection criteria for the scalar bottom quarks
search are applied, 5 events are observed and an expected background
of $5.8 \pm 1.8$ is determined. The 95\% CL excluded region in 
the $m (\tilde{\chi}_{1}^{0}) - m (\tilde{b}_{1})$ plane is shown in
the right plot of Figure \ref{fig:susy}. 
The OPAL results\cite{opal} for this decay channel are superimposed.
For $m (\tilde{\chi}_{1}^{0}) = 40$ GeV/c$^{2}$, the maximum scalar
bottom quark mass excluded is $146$ GeV/c$^{2}$ at 95\% CL.


\nonumsection{References}

\end{document}